\documentclass[a4paper,12pt]{article}
\usepackage{amsfonts,amsmath,amssymb,amscd,amsthm,latexsym,url}

\newcommand{\bi}{\bibitem}
\newcommand{\nb}{\newblock}
\newcommand{\be}[1]{\begin{equation}\label{#1}}
\newcommand{\ee}{\end{equation}}

\newtheorem{thm}{\quad Theorem}

\begin{document}
\title{Reducing SAT to 2-SAT}%
\author{Sergey Gubin}%
\thanks{Author's email: sgubin@genesyslab.com}
\begin{abstract}
Description of a polynomial time reduction of SAT to 2-SAT of polynomial size.
\end{abstract}
\maketitle

\section{Introduction}
Among all dimensions, 2-SAT possesses many special properties unique in the sense of computational complexity \cite{cook1, cook, karp, garry, lane}. But in light of works \cite{gubin3, gubin1, gubin2, gubin} a problem arose: either those properties are accidental or there are polynomial time reductions of SAT to 2-SAT of polynomial size. This article describes one such reduction.

\section{Presenting SAT with XOR}
In \cite{gubin3} was described one of the ways to present SAT with a conjunction of XOR. Let us summarize it.
\newline\indent
Let Boolean formula $f$ define a given SAT instance:
\begin{equation}
f = c_{1} \wedge c_{2} \wedge \ldots \wedge c_{m}.
\end{equation}
Clauses $c_{i}$ are disjunctions of literals:
\[
c_{i} = L_{i1} \vee L_{i2} \vee \ldots \vee L_{in_{i}}, ~ i = 1,2,\ldots,m
\]
- where $n_{i}$ is the number of literals in clause $c_{i}$; and $L_{ij}$ are the literals. Using distributive laws, formula (1) can be rewritten in disjunctive form:
\[
f = d_{1} \vee d_{2} \vee \ldots d_{p}, ~ p = n_{1}n_{2} \ldots n_{m}.
\]
Clauses $d_{k}$ in this presentation are conjunctions of $m$ literals - one literal from each clause $c_{i}, ~ i = 1,2,\ldots,m$:
\begin{equation}
d_{k} = L_{1k_{1}} \wedge L_{2k_{2}} \wedge \ldots \wedge L_{mk_{m}}, ~ k = 1,2,\ldots,p.
\end{equation}
It is obvious that formula (1) is satisfiable iff there are clauses without complimentary literals amongst conjunctive clauses (2). Disjunction of all those clauses is the disjunctive normal form of formula (1). Thus, formula (1) is satisfiable iff there are members in its disjunctive normal form.
\newline\indent
There is a generator for conjunctive clauses (2):
\begin{equation}
g = \bigwedge_{i=1}^{m} ~ (\xi_{i1} \oplus \xi_{i2} \oplus \ldots \oplus \xi_{in_{i}}) = true,
\end{equation}
- where Boolean variable $\xi_{\mu\nu}$ indicates whether literal $L_{\mu\nu}$ participates in conjunction (2). Solutions of equation (3) generate conjunctive clauses (2). Let's call the variables $\xi$ the indicators. To select from all solutions of equation (3) those without complimentary clauses, let's use another Boolean equation.
\newline\indent
For each of the combination of clauses $(c_{i},c_{j}), ~ 1 \leq i < j \leq m$, let's build a set of all couples of literals participating in the clauses:
\[
A_{ij} = \{ ~ (L_{i\mu},L_{j\nu}) ~ | ~ c_{i} = L_{i\mu} \vee \ldots; ~ c_{j} = L_{j\nu} \vee \ldots ~ \}.
\]
Let $B_{ij}$ be a set of such couples of indicators $(\xi_{i\mu},\xi_{j\nu})$, that the literals they present are complimentary:
\[
B_{ij} = \{ ~ (\xi_{i\mu},\xi_{j\nu}) ~ | ~ (L_{i\mu},L_{j\nu})\in A_{ij}, ~ L_{i\mu} = \bar{L}_{j\nu} ~ \}.
\]
There are $C_{m}^{2}$ sets $B_{ij}, ~ 1 \leq i < j \leq m$, and
\[
|B_{ij}| \leq \min\{n_{i},n_{j}\}.
\]
Let's mention that some of the sets can be empty. Then, the following equation will select from all solutions of equation (3) those without complimentary clauses:
\begin{equation}
h = \bigwedge_{1 \leq i < j \leq m} ~ \bigwedge_{(\xi,\zeta) \in B_{ij}} ~ (\bar{\xi} \vee \bar{\zeta}) = true.
\end{equation}
Due to the above estimations of the number of sets $B_{ij}$ and of their sizes, the number of clauses in formula (4) is 
\[
n = O(t_{2}m^{2}), 
\]
- where $t_{2}$ is the second number in the row of clauses' sizes sorted by value:
\[
t_{1} = \max\{n_{1},n_{2},\ldots,n_{m}\},~t_{2} = \max_{i < j}\min\{n_{i},n_{j}\}, ~ \ldots
\]
\indent
Because satisfiability of formula (1) means that the disjunctive normal form of formula (1) has conjunctive clauses, formula (1) is satisfiable iff the following formula/equation is satisfiable:
\begin{equation}
g \wedge h = true.
\end{equation}
\indent
The reasons for replacing formula (1) with formula (5) are explained in \cite{gubin3}. The number of $true$-strings in truth-tables of XOR clauses of formula (3) is linear over initial input. The number of $true$-strings in truth-tables of disjunctive clauses of formula (4) is just 3. The number of all clauses in (5) is cubic over initial input. It can be estimated as 
\[
m + n = O(t_{2}m^{2}).
\]
Thus, application of the simplified compatibility matrices method \cite{gubin3} to equation (5) will produce a polynomial time algorithm for SAT. But let's return to the reduction.

\section{SAT vs. 2-SAT}
Let's apply the simplified method of compatibility matrices \cite{gubin3} to equation (5). The method consists of sequential Boolean transformations of compatibility matrices of equation (5). Let's mention that after $m$ iterations, due to the allocation of formula (4) at the end of formula (5), there will only be compatibility matrices of equation (4) left in play. They will be grouped in an upper triangular box matrix
\begin{equation}
S = (F_{m+\mu,m+\nu})_{1 \leq \mu < \nu \leq n}.
\end{equation}
The matrix is displayed below:
\[
\begin{array}{|c|c|c|c|}
\hline
F_{m+1,m+2}&F_{m+1,m+3}&\ldots&F_{m+1,m+n} \\
\hline
 &F_{m+2,m+3}&\ldots&F_{m+2,m+n} \\
\hline
&&\ddots&\vdots \\
\hline
&&&F_{m+n-1,m+n} \\
\hline
\end{array}
\]
\indent
If there are no complimentary literals in different clauses of formula (1), then formula (4) is just missing. The size of matrix (6) is $0 \times 0$. In this case, formula (1) is reducible to 1-SAT instance
\[
\omega_{1} \wedge \omega_{2} \wedge \ldots \wedge \omega_{m},
\]
- where
\[
\omega_{i} = \xi_{i1} \oplus \xi_{i2} \oplus \ldots \oplus \xi_{in_{i}}, ~ i = 1,2,\ldots,m.
\]
This singularity belongs to the set of all 2-SAT instances.
\newline\indent
If, during the first $m$ iterations, a pattern of unsatisfiability arises (one of the compatibility matrices becomes filled with $false$ entirely), then formulas (5) and (1) are both unsatisfiable \cite{gubin3}. This case may be thought of as a case of formula (1) being reduced to an unsatisfiable formula
\[
false.
\]
Let's include this singularity in the set of all 2-SAT instances.
\newline\indent
Otherwise, boxes $F_{m+\mu, m+\nu}$ in matrix (6) are what is left of the compatibility matrices of equation (4) after the first $m$ iterations of the method. 
\newline\indent
Due to their construction \cite{gubin3}, the boxes are $3 \times 3$ matrices:
\begin{equation}
F_{m+\mu, m+\nu} = (x_{ij})_{3 \times 3}, ~ 1 \leq \mu < \nu \leq n
\end{equation}
- where $x_{ij} \in \{false, true \}$. The number of boxes is $C_{n}^{2}$. Thus, the number of all elements in matrix (6) is
\[
e = 9C_{n}^{2} = O(t_{2}^{2}m^{4}).
\]
Let's enumerate the elements arbitrarily:
\[
y_{1}, ~ y_{2},\ldots,y_{e}.
\]
Then, distribution of $true$/$false$ in matrix (6) can be described with a 1-SAT formula/equation
\begin{equation}
w = \eta_{1} \wedge \eta_{2} \ldots \wedge \eta_{e} = true,
\end{equation}
- where $\eta_{i}$ are literals over a set of Boolean variables
\[
\{~b_{1}, ~ b_{2}, \ldots, b_{e}~\}.
\]
The literals are
\[
\eta_{i} = \left \{ \begin{array}{ll}
b_{i}, & y_{i} = true \\
\bar{b}_{i}, & y_{i} = false
\end{array}, \right.
~ i = 1,2,\ldots,e.
\]
\indent
Let's take the following 2-SAT instance:
\begin{equation}
h \wedge w.
\end{equation}
Box matrix (6) is an initialization of the modified method of compatibility matrices \cite{gubin3} for formula (9): compatibility matrices of formula (4) are depleted to satisfy equation (8). Thus, continuation of the simplified method of compatibility matrices for equation (5) from its Step $m+1$ to its finish is an application of the modified method of compatibility matrices to system (9) from its Step 1 to its finish \cite{gubin3}. After $n-2$ iterations, both methods must result with the same version of satisfiability of formula (1). Thus, formulas (5) and (1) are satisfiable iff 2-SAT formula (9) is satisfiable. The number of clauses in formula (9) is
\[
e + n = O(t_{2}^{2}m^{4}).
\]
According to \cite{gubin3}, the time to deduce formula (9) can be safely estimated as
\[
O(t_{1}^{4}t_{2}^{4}m^{6})
\]

\section{SAT vs. 1-SAT}
Let's take one step further. Applying to formula (1)/(5) either of the variations of the compatibility matrices method \cite{gubin3} will produce a Boolean matrix. Let it be a matrix $R$:
\[
R = (r_{ij})_{a \times b}.
\]
Size of the matrix depends on the method's variation and the order of clauses in formula (1). The size can be changed if permute the clauses and repeat the method \cite{gubin3}. The formula (1) is satisfiable iff matrix $R$ contains $true$-elements \cite{gubin3} (elements which are $true$). The existence/absence of the $true$-elements is the only invariant.
\newline\indent
If formula (1) is unsatisfiable, then that formula is reducible to formula ``$false$''. Otherwise, formula (1) is reducible to a 1-SAT instance.
\proof
Let's enumerate elements of matrix $R$ in arbitrarily order:
\[
z_{1}, z_{2}, \ldots, z_{ab}.
\]
Let $B$ be a set of $t = ab$ Boolean variables:
\[
B = \{~b_{i} \in \{false,true\}~|~i = 1,2,\ldots,t~\}.
\]
Then the following 1-SAT formula describes distribution of $true/false$ in matrix $R$:
\begin{equation}
\theta_{1} \wedge \theta_{2} \wedge \ldots \wedge \theta_{t},
\end{equation}
- where literals $\theta_{i}$ are
\[
\theta_{i} = \left \{ \begin{array}{ll}
b_{i}, & z_{i} = true \\
\bar{b}_{i}, & z_{i} = false
\end{array}, \right.
~ i = 1,2,\ldots,t.
\]
Thus, the compatibility matrices method reduces satisfiable formula (1) to 1-SAT formula (10).
\endproof
\indent
In its turn, formula (10) can be rewritten as SAT of any dimension by appropriate substitution of variables.
\newline\indent
If use the simplified method of compatibility matrices, then matrix $R$ is a $3 \times 3$ Boolean matrix \cite{gubin3}. Let there be two clauses shorter than $3$ in formula (1). Let's permute all clauses and make those shortest clauses to be the last ones in formula (1). Then, result of the modified method \cite{gubin3} will be a matrix $R$ of size less than $3 \times 3$. That proves the following theorem.
\begin{thm}
Any SAT instance is reducible to a 1-SAT instance with $9$ variables or less. A SAT instance is unsatisfiable iff its 1-SAT presentation is ``$false$'' - there is not any variables in its 1-SAT presentation.
\end{thm}

\section{Conclusions}
Formula (1) may be thought of as a ``Business Requirements''. And any appropriate computer program may be thought of as a solution of the SAT instance. Then, theorem 1 can be an explanation of the remarkable efficiency of the ``natural programs''. From this point of view, the iterations of the method of compatibility matrices may be thought of as a learning/modeling of the business domain. In the artificial programming, the calculation of the compatibility matrices - a virtual business domain - could be a conclusion of the stage ``Business Requirements Analysis/Mathematical Modeling''. That would improve the programs' performance. The resulting compatibility matrices may be thought of as a fussy logic's tables of rules for the domain. 
\newline\indent
The whole solution of formula (1) can be achieved, with one of the following approaches, for example. ANN approach is the applying of the compatibility matrices method backward, starting from matrix $R$. An example of that can be found in \cite{gubin2}. DTM approach is the looping trough of the following three steps: selection of any $true$-element from matrix $R$; substitution of the appropriate $true$-assignments in formula (1); and repeating of the compatibility matrices method. The last method is an implication of the self-reducibility property of SAT \cite{lane}. 
\newline\indent
In certain sense, theorem 1 may be seen as an answer to the Feasibility Thesis \cite{cook}.

\end{document}